\RequirePackage{etex}

\documentclass[]{aa}

\usepackage{graphicx}
\usepackage[varg]{txfonts}
\usepackage{physics}
\usepackage{amsmath, amssymb}
\usepackage{dsfont}
\usepackage{bbm}

\usepackage{ulem}
\usepackage{xpatch}
\usepackage{hyperref}
\usepackage{todonotes}
\usepackage{import}
\usepackage{makeidx}
\usepackage{changes}
\usepackage{placeins}
\usepackage{soul}
\usepackage[flushleft]{threeparttable}

\newcommand{\tab}[1]{Table~\ref{#1}}
\newcommand{\fig}[1]{Fig.~\ref{#1}}

\newcommand{\equ}[1]{Eq.~(\ref{#1})}
\newcommand{\equo}[1]{Eq.~\ref{#1}}

\newcommand{\Msolpyr}{\mathrm{M_\odot~yr^{-1}}}
\newcommand{\Msol}{\mathrm{M_\odot}}

\newcommand{\colout}[1]{\bgroup\markoverwith{\textcolor{#1}{\rule[.5ex]{2pt}{0.4pt}}}\ULon}

\makeatletter
\renewcommand*\aa@pageof{, page \thepage{} of \pageref*{LastPage}}

\makeatother

\renewcommand{\arraystretch}{1.5}

\makeindex

\begin{document}

\titlerunning{The effect of variable stellar magnetic fields on the spin state of T Tauri stars}

\authorrunning{L.~Gehrig et al.}
\title{The effect of variable stellar magnetic fields on the spin state of T Tauri stars} 

\author{
 Lukas~Gehrig and
 Daniel~Steiner
}

\institute{
 Department of Astrophysics, University of Vienna,
 Türkenschanzstrasse 17, 1180 Vienna, Austria
}

\date{Received ....; accepted ....}

\abstract
{
Understanding the stellar spin evolution of young stars is crucial for understanding the evolution of protoplanetary disks and, consequently, the formation of exoplanets.
According to stellar spin models, T Tauri stars should evolve toward a spin equilibrium in which the external spin-down torque balances both the external spin-up (accretion) torque and the spin-up due to stellar contraction. A useful reference point along the way to this equilibrium is the "zero-torque state" (ZTS), at which only the external torques cancel out. Recent observations, however, have shown that the spin state of a considerable number of stars is shifted out of the spin equilibrium and the ZTS.
We investigate the effects of variable stellar magnetic fields on the stellar spin state of T~Tauri stars. 
The long-term decrease ($0.1- 2.0$~Myr) in the stellar field strength due to the formation of a radiative core and short-term stellar magnetic cycles ($10^1-10^4$~years) are taken into consideration.
Both scenarios are studied using an implicit disk evolution code capable of modeling the magnetic star--disk interaction in the innermost disk and the stellar spin evolution self-consistently over long timescales.
Temporal variations in the stellar magnetic field can significantly affect the stellar spin state of T~Tauri stars.
The strength of the effect on the stellar spin state depends on the relation between the timescale of the changing magnetic field, the spin-up timescale, and the viscous timescale of the accretion disk.
A developing radiative core on a timescale shorter than the spin-up timescale has a strong effect on the spin state.
Stellar magnetic cycles on timescales shorter than the viscous timescale of the inner disk have a weaker effect on the stellar spin state due to a slow back-reaction of the accretion disk.
Our results can explain (at least) a part of the stars that are observed out of both states.
Further theoretical and observational work is needed to connect accretion, stellar rotation, and magnetic properties in T~Tauri stars.
}


\maketitle

\section{Introduction}
\label{sec:intro}

To understand the formation and evolution of exoplanets, we require insights into the early processes in classical T Tauri stars (CTTSs; $M_\star \lesssim 2~\mathrm{M_\odot}$) and their surrounding protoplanetary disk (PPD) that shape the cradle of young systems.
The stellar spin distribution observed in CTTSs poses an ongoing challenge.
Observations show rotation periods ranging between $\sim1$ and $10$ days \citep[e.g.,][]{Rebull2006, herbst02, Cody2010, Venuti2017, Rebull2020, Serna2021, Smith2023}.
During the CTTS phase, young stars actively accrete material from their PPDs, increasing their angular momentum (AM), and the stars contract, which should lead to a spin-up \citep[e.g.,][]{Rebull2002}.
The rotation distributions of young stars, however, remain relatively constant over the expected PPD lifetime of $\lesssim 10$~Myr \citep[e.g.,][]{Smith2023} and show a spin-up after most PPDs have dissipated \citep[e.g.,][]{Gallet13}.

To prevent the spin-up of CTTSs, the magnetic star--disk interaction presumably removes stellar AM \citep[e.g.,][]{Koenigl91}.
In early studies, the disk was assumed to be "locked" to the star, removing stellar AM \citep[e.g.,][]{Ghosh79, Shu94}.
This "disk-locking" mechanism, however, has since been found to be too ineffective to explain the observed spin distributions \citep[e.g.,][]{Agapitou2000, Rebull2001, matt05, Herbst2007, Zanni2009}.
In more recent models \citep[e.g.,][]{Romanova2018, Gallet19, Ireland21, Takasao2022, Takasao2025, Zhu2025}, stellar AM is removed by outflows in the form of magnetospheric ejections \citep[MEs;][]{Zanni13}, conical disk winds \citep[CDWs; e.g.,][]{Romanova04, Lii2014, Romanova2018, Takasao2025}, and accretion-powered stellar winds \citep[APSWs; e.g.,][]{Matt05APSW, Matt2012APSW, Finnley18}.
The amount of AM that can be removed from the star depends on the relation between the inner disk (or truncation) radius, $r_\mathrm{t}$, and the corotation radius, $R_\mathrm{co} = (GM_\star P_\star^2 / 4 \pi^2)^{1/3}$, where  $M_\star$ is the stellar mass and $P_\star$ is the stellar rotation period \citep[e.g.,][]{Romanova09, Gallet19, Ireland2022, Takasao2022, Takasao2025}.
This relation is often parameterized with the fastness parameter, $\omega_\mathrm{s}=(r_\mathrm{t}/R_\mathrm{co})^{3/2}$ \citep[][]{Ghosh2007}.
The magnetic field strength of a young star plays a crucial role in its spin evolution, as it controls the net amount of AM that different spin-down mechanisms can carry away \citep[e.g.,][]{Zanni13, Gallet19, Ireland2022, Takasao2025} and the location of the truncation radius, $r_\mathrm{t}$ \citep[e.g.,][]{hartmann16}, influencing $\omega_\mathrm{s}$.

With an increasing value of $\omega_\mathrm{s}$, the spin-down mechanisms can carry away more AM from the star \citep[e.g.,][]{Ireland2022, Takasao2025}, resulting in a stabilizing effect on the stellar spin.
For a star with a given inner disk radius, $r_\mathrm{t}$, a fast rotation period leads to a small $R_\mathrm{co}$ and a large $\omega_\mathrm{s}$.
Consequently, the spin-down mechanisms are removing stellar AM efficiently, causing the star to spin down and decreasing the $\omega_\mathrm{s}$.
Similarly, a star that spins slowly,  equivalent to a small value of $\omega_\mathrm{s}$, spins up.

A further consequence is that the external torques on the star tend toward a balance, the so-called zero-torque state (ZTS) at which external spin-up (e.g., accretion torques) and external spin-down contributions cancel each other out \citep[e.g.,][]{Pantolmos20, Ireland21, Ireland2022, Takasao2025, Zhu2025}. 
A star in a ZTS, however, is still subject to spin-up from contraction. Stars therefore evolve toward a spin-equilibrium state, in which external spin-down balances both the accretion torque and the contraction-driven spin-up. This equilibrium is reached at a fastness parameter larger than that of the ZTS, with the exact value depending on the stellar torque model and the contraction timescale. 
This distinction does not affect the conclusions of the present work, as the fastness parameters of interest, both in our models and in recent observations, are smaller than those of either state.

Depending on the stellar parameters and the presence of an APSW, ZTS occurs at $0.3 \lesssim \omega_\mathrm{s} \lesssim 1.0$ \citep[e.g.,][]{Gehrig2025Letter, Zhu2025}. 
The evolutionary timescale on which the star reaches the ZTS, $\tau_\mathrm{J}$, is expected to be on the order of a few million years \citep[e.g.,][]{Lii2014, Ireland21, Serna2024}.
Based on theoretical models,  CTTSs with ages of $\gtrsim 2$~Myr are expected to spin at (or close to) the ZTS.
However, recent measurements of $R_\mathrm{i}$ and $R_\mathrm{co}$ have shown that a considerable number of CTTSs are in a spin-up state with values of $\omega_\mathrm{s} < 0.3$ \citep[e.g.,][]{Thanathibodee2023, Pittman2025rot}, which is smaller than the expected values of the ZTS \citep[e.g.,][]{Gehrig2025Letter}.
While some stars might be too young to have reached their ZTS, these observations challenge the assumption that stars evolve toward a zero torque condition.

In this work, we studied the effects of temporal variations in the stellar dipole field strength, $B_\star$, on the stellar spin evolution during the T Tauri phase of young stars.
The stellar dipole field strength is an important parameter as it influences the stellar spin model \citep[e.g.,][]{Gallet19, Ireland2022} as well as the location of the inner disk radius \citep[e.g.,][]{bessolaz08, hartmann16}.
We considered two different timescales on which $B_\star$ can change.
First, the timescale $\tau_\mathrm{B}$ over which CTTSs undergo magnetic topology changes when they develop a radiative core \citep[e.g.,][]{Gregory2012, Lavail19}.
Fully convective stars tend to have magnetic field configurations that are dominated by a low-order dipole with field strengths ($B_\star$) of $\sim 1$~kG.
As a star with a sufficient mass develops a radiative core, the magnetic field becomes more complex, with dominant high-order field components. 
The dipole component for these stars is up to an order of magnitude weaker: $B_\star \sim 0.1$~kG \citep[e.g.,][]{Gregory2012}.
The time in which CTTSs develop a radiative core depends on the stellar mass.
As stated in \cite{Gregory2012}, a $1~\mathrm{M_\odot}$ star develops a radiative core after approximately 2~Myr and a $0.7~\mathrm{M_\odot}$ star after 6~Myr, within the expected lifetime of the PPDs.
The timescale on which the magnetic field adjusts to the development of the radiative core is estimated to be $\sim 0.1-1.0$~Myr \citep[e.g.,][]{Gregory2012, Villebrun2019}.

Second, we examined shorter stellar magnetic cycles, which affect $B_\star$.
One mechanism that affects the stellar magnetic field is the reversal of the stellar magnetic field polarity, which can lead to a distinct change of the dipole field \citep[like the 11-year solar cycle; e.g.,][]{DeRosa2012, Lehtinen2022, Finociety2023, Gaidos2024}.
As shown in \cite{Johnstone14}, the stellar dipole field strength can vary significantly (from 280~G to 970~G) within a time span of several years.
However, we do not know if regular, well-defined stellar cycles such as the solar cycle exist in T~Tauri stars.
The reconstruction of the solar magnetic activity over the past few millennia has also shown variations on longer timescales \citep[e.g.,][]{Usoskin2021}.
Due to the limited data available, the applicability of such variations to T~Tauri stars requires further investigation.
In this work, we assumed that magnetic cycles affect the stellar dipole field in T~Tauri stars on timescales of $\sim 10$~years.
As the stellar dipole field changes over time, the inner disk and corotation radius, and thus $\omega_\mathrm{s}$, are affected.
Starting with a star rotating in its ZTS, we modeled how a time-dependent magnetic field affects the stellar spin evolution.

\section{Model description}
\label{sec:model}

To model the structure and long-term evolution of PPDs, we used the implicit hydrodynamic TAPIR code \citep[e.g.,][]{ragossnig20TAPIR, Steiner21}.
The equations of hydrodynamics are combined with the description of the turbulent viscosity presented in \cite{Shakura1973}.
We used a layered viscosity model \citep[][]{gammie96} featuring a low turbulent or dead zone (DZ) with a viscosity parameter $\alpha_\mathrm{DZ}$ and a turbulent layer with $\alpha_\mathrm{turb}\gg\alpha_\mathrm{DZ}$.
Using the TAPIR model, we could incorporate the effects of stellar magnetic torques resulting from the star--disk interaction, which cause a deviation from the Keplerian velocity within the disk, as the diffusion approach \citep[e.g.,][]{pringle81, Armitage01} is not employed.
Furthermore, the model uses an adaptive grid \citep[e.g.,][]{Dorfi1987}, allowing the inner and outer disk boundaries to move according to variable stellar and disk parameters \citep[e.g., the accretion rate or stellar contraction][]{Steiner21, Gehrig2022}.
In its further development, the model has been enhanced with a stellar spin model \citep[][]{Gehrig2022} and an X-ray photo-evaporative model \citep[][]{Cecil2024}, in which the stellar X-ray luminosity controls the photo-evaporative mass-loss rate, $L_\mathrm{X}$.
As a result, the TAPIR model can combine the effects of the magnetic star--disk interaction in the innermost disk, stellar rotation, and photoevaporation self-consistently, making it well suited for the aims of this study.
We note that magnetohydrodynamic-driven disk winds are not considered in the current version of the model.
The effects of such winds on the results are discussed in Sect. \ref{sec:timescales}.
We used the model as described in \cite{Cecil2024}.
A detailed description of the equations solved in the model and the underlying assumptions is summarized there.
Here, we highlight specific aspects of the model and list the used parameters.

\subsection{Calculation of the truncation radius}

The inner disk or truncation radius, $r_\mathrm{t}$, is calculated by equating the magnetic pressure from the stellar magnetic field, $P_\mathrm{mag}$, and the pressure of the disk material containing the ram pressure of the accreting material, $P_\mathrm{ram}$, and the gas pressure, $P_\mathrm{gas}$ \citep[e.g.,][]{Koldoba02,Romanova02,bessolaz08},
\begin{equation}\label{eq:r_trunc}
    P_\mathrm{magn}(r_\mathrm{t}) = \text{max}\left[P_\mathrm{ram}(r_\mathrm{t}), P_\mathrm{gas}(r_\mathrm{t})\right]\;.
\end{equation}
For the case $P_\mathrm{ram} \geq P_\mathrm{gas}$, we followed \cite{hartmann16},
\begin{align}
    r_{\mathrm{t}}(P_\mathrm{ram} \geq P_\mathrm{gas}) \approx 18 \, \xi \, R_\odot & \, \left(\frac{B_\star}{10^3 \, G}\right)^{4/7}  \left(\frac{R_\star}{2 \, R_\odot}\right)^{12/7} \left(\frac{M_\star}{0.5 \, M_\odot}\right)^{-1/7} \nonumber \\ 
    & \left(\frac{\dot M_\star}{10^{-8} \, M_\odot / \mathrm{yr}}\right)^{-2/7} \label{eq:truncation_radius} \;, 
\end{align}
where $B_\star$ is the stellar magnetic field strength, $R_\star$ the stellar radius, $M_\star$ the stellar mass, and $\Dot{M}_\star$ the accretion rate on the star, and the correction factor ($\xi$) is $0.7$ \citep[][]{hartmann16}{}{}.
For $P_\mathrm{ram} < P_\mathrm{gas}$, the truncation radius follows from the condition 
\begin{equation}
    P_\mathrm{gas}(r_\mathrm{t}) = P_\mathrm{magn}(r_\mathrm{t}) = \frac{B_\star^2R_\star^6}{8 \pi r_\mathrm{t}^6}~.
\end{equation}

\subsection{Stellar spin model}

To calculate the stellar spin evolution, we included an external torque acting on the star,  $\Gamma_\mathrm{ext} = \Dot{J}_\star$, which equals the temporal derivative of the stellar AM, and the internal stellar evolution \citep[e.g.,][]{Gallet19}.
We assumed that the star rotates as a rigid body, $J_\star = \Omega_\star I_\star$, with the stellar moment of inertia, $I_\star = 0.2 M_\star R_\star^2$ \citep[e.g.,][]{Armitage96, Matt10}\footnote{
A developing radiative core in a young star eventually decouples from the convective envelope. 
However, the timescale at which the core rotation diverts significantly from the envelope is assumed to be on the order of $\gtrsim 10$~Myr \citep[e.g.,][]{Gallet13}, which is older than most T~Tauri systems.
Thus, the assumption of rigid stellar rotation should be valid within the scope of this study. }.
This results in a temporal derivative of the stellar angular velocity of
\begin{equation}\label{eq:omegadot}
    \Dot{\Omega}_\star = \frac{\Gamma_\mathrm{ext}}{I_\star} - \frac{\Dot{I}_\star}{I_\star} \Omega_\star \,,
\end{equation}
with
\begin{equation}
    \frac{\Dot{I}_\star}{I_\star} = \frac{\Dot{M}_\star}{M_\star} + 2 \frac{\Dot{R}_\star}{R_\star} \, .
\end{equation}
The external torque consists of the effects of the accretion process, $\Gamma_\mathrm{acc}$, of the APSW, $\Gamma_\mathrm{APSW}$, and of the ME, $\Gamma_\mathrm{ME}$; $\Gamma_\mathrm{ext} = \Gamma_\mathrm{acc} + \Gamma_\mathrm{APSW} + \Gamma_\mathrm{ME}$ \citep[see also][]{Gallet19, Gehrig2022}.
The accreted material carries AM that is added to the star and spins it up.
An APSW removes AM from the star and causes a spin-down effect.
MEs can spin up or spin down the star depending on the value of $\omega_\mathrm{S}$.
In the model of \cite{Gallet19}, MEs spin up the star for $\omega_\mathrm{S}\leq0.7$.

Following \cite{Matt10}, the star contracts along the Hayashi track on its Kelvin-Helmholtz timescale\footnote{
This assumption is usually used to describe fully convective stars. However, the stellar radius evolution described here does not deviate significantly from stellar evolution tracks \citep[e.g.,][]{Baraffe15} within the first $\sim 10$~Myr for stellar masses of $M_\star\lesssim 1.0~\mathrm{M_\odot}$.
}
, resulting in a radius change according to
\begin{equation}\label{eq:contraction}
    \Dot{R}_\star = 2 \frac{R_\star}{M_\star} \Dot{M}_\star - \frac{28 \pi \sigma R_\star^4 T_\mathrm{eff}^4}{3 G M_\star^2} \, 
,\end{equation}
where $T_\mathrm{eff}$ is the stellar effective temperature and $\sigma$ the Stefan-Boltzmann constant.
A detailed summary of the respective equations can be found in our previous work \citep[][]{Gehrig2022} and in \cite{Gallet19}.

\subsection{Variable magnetic field}\label{sec:Bvar}

\begin{figure}
    \centering
         \resizebox{\hsize}{!}{\includegraphics{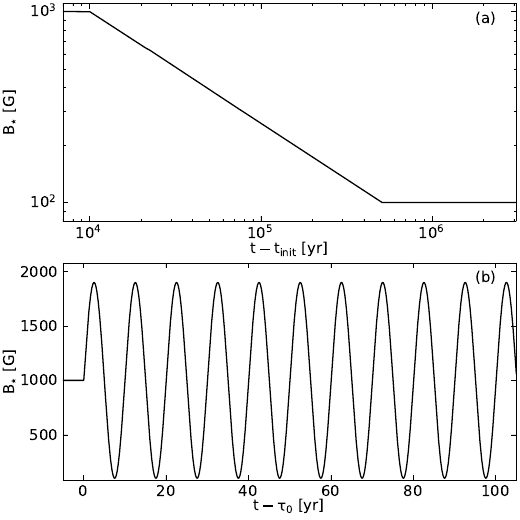}}
    \caption{Illustration of the time-dependent dipole field strength used in this work. Panel (a): Long-term evolution due to the development of a radiative core. Panel (b): Short-term variation due to stellar cycles.
    }
    \label{fig:Bvar}
\end{figure}

We studied the effects of a time-dependent magnetic field strength on two different timescales.
First, the long-term effect of a decrease in the stellar dipole field strength due to the development of a radiative core was simulated (called the "long model").
According to \cite{Gregory2012}, fully convective T~Tauri stars host magnetic fields with a dominant dipole component of $B_\star \gtrsim 1$~kG.
Stars that have developed radiative cores, on the other hand, show more complex fields with dominant higher-order field components and weaker dipole fields of $B_\star \sim 0.1$~kG.
Recent observations suggest that different dynamo mechanisms are responsible for generating the magnetic field in fully convective and partially convective stars \citep[e.g.,][]{See2025}, with notable differences particularly in the large-scale or dipole field strengths.
It is assumed that the dipole component is the dominant component controlling the star--disk interaction \citep[e.g.,][]{Finnley18}.
The transition between these two states should occur within a few tenths of a megayear \citep[e.g.,][]{Gregory2012, Villebrun2019}.
In our model, we started with a stellar dipole field of $B_\star = 1$~kG at the beginning of the simulations.
The initial stellar age, $t_\mathrm{init}$, corresponds to the development of the radiative core and depends on the stellar mass (see Sect. \ref{sec:paras} and \tab{tab:model_para}).
After a simulation time of $\tau_\mathrm{0}=10^4$~years, $B_\star$ decreases and reaches 0.1~kG after $\tau_\mathrm{B}=0.5$~Myr (see panel a of \fig{fig:Bvar}).

Second, we included short-term variations due to stellar cycles \citep[e.g.,][referred to as the "short model"]{Armitage1995}.
During a cycle period of $\tau_\mathrm{cyc}$, the stellar field strength varies according to
\begin{equation}\label{eq:cycle}
    B_\mathrm{\star}(t) = B_\mathrm{mean} + B_\mathrm{osc} \sin{\left( \frac{2 \pi t}{\tau_\mathrm{cyc}} \right) } \, ,
\end{equation}
where $B_\mathrm{mean}$ is the mean dipole field component and $B_\mathrm{osc}$ the oscillating field component.
Again, we started with the variation of $B_\star$ after $\tau_\mathrm{0}=10^4$~years.
In fact, measurements of stellar dipole fields show a strong variability of $B_\star$ over several years. 
For example, the dipole field of V2129~Oph has increased from 280~G in 2005 to 970~G in 2009 \citep[][]{Johnstone14}.
Based on the variation in photometric data of T~Tauri stars ranging on timescales of $1.5-4$~years \citep[e.g.,][]{Lin2023}, the cycle timescale can be estimated to be around $\sim 3$~years \citep[e.g.,][]{Gaidos2024}.
Furthermore, models of cool dwarfs \citep[e.g.,][]{Bice2023} and solar-like stars \citep[e.g.,][]{Strugarek2018} suggest plausible cycle timescales of $\sim1$~year and $\sim 30$~years, respectively.
Thus, we chose $\tau_\mathrm{cyc}=10$~years \citep[following][]{Armitage1995}, $B_\mathrm{mean} = 1.0$~kG, and $B_\mathrm{osc}=0.9$~kG.
An illustration of $B_\star(t)$ is given in \fig{fig:Bvar}b.
Short-term variations of the stellar magnetic field of T~Tauri stars are associated with a polarity reversal of the magnetic field \citep[similar to the 11-year solar cycle; e.g.,][]{DeRosa2012, Finociety2023, Gaidos2024}.
We note that the adopted form of the short-term magnetic cycle (\equo{eq:cycle}) corresponds to an "effective" or absolute dipole field component acting on the disk.
During a polarity reversal, the dipole field would decrease toward zero and change sign \citep[as recently reported in][]{Donati2026}.
For a dipole field close to zero, field components of higher order (quadrupole or octopole fields) start to affect the disk, resulting in an effective, nonzero component even at the polarity inversion \citep[e.g.,][]{Finnley18}. 
In addition, we assumed that the alignment between the dipole field and the disk's rotational axis is fixed.
We discuss additional effects due to the changing alignment between the magnetic and rotational axes in Sect. \ref{sec:reversal}.
Further, the sinusoidal form of the cycle \citep[following][]{Armitage1995} is an approximation.
Observed magnetic field intensities in young stars show rather abrupt changes in the dipole component during possible polarity reversals \citep[e.g.,][]{Finociety2023}.
Although the exact form of the cycle remains unclear, the results of this work should not be significantly affected by a different cycle form.
We chose the sinusoidal representation for its numerical stability.

\subsection{List of parameters}\label{sec:paras}

Here, we provide an overview (see \tab{tab:model_para}) of the parameters used in this work.
We modeled three different stellar masses. 
To study the effects of a developing radiative core, we chose $M_\star = 1.0$ and $0.7~\mathrm{M_\odot}$.
According to stellar evolution models, a solar mass star develops a radiative core shortly after $2$~Myr and a $0.7~\mathrm{M_\odot}$ star at around 5~Myr \citep[e.g.,][]{Siess00, Tognelli11}, which is the assumed stellar age at the beginning of the respective simulations, $t_\mathrm{init}$.
Magnetic cycles are also modeled for these stars, and, in addition, for a $0.3~\mathrm{M_\odot}$ star with an initial age of 2~Myr.
The initial stellar radii for the respective stars are based on isochrone models from \cite{Baraffe15}, and the initial accretion rates are chosen to be within the observed range reported in \cite{Betti2023}.
The viscosity parameters are taken from \cite{Cecil2024}; $\alpha_\mathrm{turb} = 0.02$ and $\alpha_\mathrm{DZ} = 0.0004$.
The initial rotation periods correspond to the respective ZTS ($\Gamma_\mathrm{ext} \approx 0$) assuming an APSW efficiency of $W=1$~\% and $B_\mathrm{\star,~init} = 1.0$~kG.
The fastness parameter of the ZTS resulting from these input values is $\omega_\mathrm{S} \approx 0.72$, which matches the results of recent simulations \citep[][]{Zhu2025} and population models \citep[][]{Gehrig2025Letter}.
We initialized at ZTS rather than at spin equilibrium to isolate the effect of the variable magnetic field.
We note that the star will spin up from stellar contraction in this configuration.
Finally, we needed to choose the stellar X-ray luminosity, which is responsible for the photo-evaporative mass-loss.
Recent developments in photo-evaporative models have shown lower mass-loss rates (by up to one order of magnitude) due to enhanced cooling \citep[e.g.,][]{Sellek2024} compared to previous models \citep[e.g.,][]{Ercolano2021}. 
To get a mass-loss rate in our models that is comparable to the results presented in \cite{Sellek2024}, we reduced the X-ray luminosities from \cite{Gudel2007} by a factor of 20.

\begin{table}[ht]
\centering
\caption{Initial model parameters used in this work.}              
\begin{tabular}{l | c c c}         
\hline\hline 

$M_{\star,~\mathrm{init}}$ [$\mathrm{M_\odot}$] & 1.0 & 0.7 & 0.3 \\
$t_\mathrm{init}$~[Myr] & 2.0 & 5.0 & 2.0 \\
$R_\mathrm{\star,~init}$ [$R_\odot$] & 1.917 &  1.243 & 1.238 \\
$T_\mathrm{eff}$ [K] & 4350 & 3960  & 3428 \\
$L_\mathrm{X}~\mathrm{[erg/s]}$ & $1\times 10^{29}$ & $5\times 10^{28}$ &  $1\times 10^{28}$ \\
$\Dot{M}_\mathrm{\star,~init}$ [$\mathrm{M_\odot/yr}$] & $10^{-8}$ &$3 \times 10^{-9}$& $2\times10^{-9}$  \\
$P_\mathrm{\star,~init}$ [days] & 4.2 & 3.0 & 6.0 \\

\hline\hline                                            
\end{tabular}
\label{tab:model_para}  
\end{table}

\section{Results}
\label{sec:results}

\begin{figure}
    \centering
         \resizebox{\hsize}{!}{\includegraphics{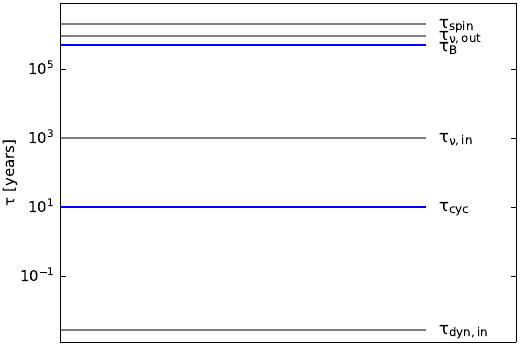}}
    \caption{Evolution timescales of the star--disk system relevant for this work. 
    The timescales on which the magnetic field is assumed to change are marked in blue.
    }
    \label{fig:timescales}
\end{figure}

Before analyzing the evolution of the long and short models, we wanted to compare the relevant evolution timescales for the star--disk system (see \fig{fig:timescales}).
In the inner disk, the truncation radius adjusts on the order of days, equivalent to the dynamic timescale, $\tau_\mathrm{dyn,in}\sim$~days \citep[e.g.,][]{Pittman2025rot}, and the accretion rate evolves on the viscous timescale, $\tau_\mathrm{\nu,in} \sim 1000$~years.
The viscous timescale in the outer disk, $\tau_\mathrm{\nu,out}\sim 9\times 10^5$~years, is a measure for the long-term evolution of the disk accretion rate and the disk lifetime.
The stellar spin evolves on a timescale $\tau_{\rm spin} \equiv \Omega_\star/\dot{\Omega}_\star$ that combines the effect of the external torques and of stellar contraction. Neglecting the slow change of the stellar mass, Eqs.~(\ref{eq:omegadot})--(\ref{eq:contraction}) yield
\begin{equation}
\tau_{\rm spin} = \frac{\tau_\mathrm{J}\,\tau_{\rm KH}}{\tau_{\rm KH} + 2\,\tau_\mathrm{J}}\,,
\label{eq:tauspin}
\end{equation}
where $\tau_\mathrm{J} = J_\star/|\dot{J}_\star|$ is the timescale associated with the external (accretion + spin-down) torques and $\tau_{\rm KH} = R_\star/|\dot{R}_\star|$ is the Kelvin-Helmholtz timescale set by the second term on the right-hand side of Eq.~(\ref{eq:contraction}). The external-torque timescale $\tau_J$ is not constant: it can reach $\sim (1-2) \times 10^6$~yr in the spin-down regime at large fastness parameters \citep[e.g.,][]{Ireland2022}, but becomes considerably shorter in an accretion-dominated spin-up regime. For our initial conditions, the system starts in a ZTS ($\tau_J \to \infty$), and $\tau_{\rm spin}$ is set by contraction, $\tau_{\rm spin} \approx \tau_{\rm KH}/2$ of approximately a few megayears.
Once the simulation starts and the value of $B_\star$ changes, the star moves out of the ZTS, and $\tau_\mathrm{J}$ becomes variable.
During our simulations, $\tau_\mathrm{spin}$ stays at values of $\gtrsim \tau_\mathrm{\nu,out}$.
Depending on the relation between these timescales and $\tau_\mathrm{B}=5\times10^5$~years and $\tau_\mathrm{cyc}=10$~years, we expect different reactions of the disk on the variation of $B_\star$.
Since $\tau_\mathrm{B}\gg \tau_\mathrm{\nu,in}\gg\tau_\mathrm{dyn,in}$, we expect that the inner disk quickly adjusts to magnetic variations due to a growing radiative core. 
The long-term evolution of the disk and the stellar rotation period, on the other hand, should only slowly adjust to the new magnetic field strength as $\tau_\mathrm{B} < \tau_\mathrm{\nu,out}<\tau_\mathrm{spin}$.
In the case of magnetic cycles, the viscous timescales in the disk and the stellar spin evolution timescale are slow to react to the short-term changes in $B_\star$.
The position of the truncation radius, however, should adjust to the variations quickly.
Thus, we can expect a nontrivial reaction in the inner disk due to the short magnetic cycles.
Overall, the stellar spin evolution (and the position of the corotation radius) is slow to react to changes in the magnetic field, and the truncation radius adjusts quickly.
Thus, we expect the stellar spin state, defined by the fastness parameter $\omega_\mathrm{s}$, to be directly affected by a variable $B_\star$.

\begin{figure}
    \centering
         \resizebox{\hsize}{!}{\includegraphics{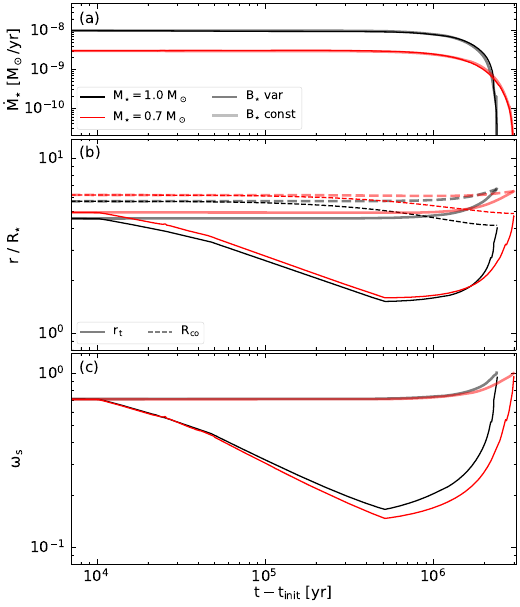}}
    \caption{Evolution of the long model compared to simulations with a constant magnetic field strength. 
    The initial stellar ages of the 1.0 and 0.7~$M_\odot$ models are $t_\mathrm{init}=$~2.0 and 5.0~Myr (see \tab{tab:model_para}), respectively.
    Panel (a): Accretion rate.
    Panel (b): Truncation radius, $r_\mathrm{t}$, and the corotation radius, $R_\mathrm{co}$, in units of the stellar radius.
    Panel (c): Fastness parameter, $\omega_\mathrm{S}$.
    }
    \label{fig:long}
\end{figure}

We started our simulations with a star in the ZTS with a fastness parameter of $\omega_\mathrm{S} \approx 0.72$. 
After $\tau_\mathrm{0}=10^4$~years, the temporal variation of $B_\star$ begins.
In \fig{fig:long}, the results for the long model are shown for stellar masses of 1.0 and $0.7~\mathrm{M_\odot}$.
For comparison, we include simulations in which the magnetic field remains constant.
With a decrease in $B_\star$ after $\tau_\mathrm{0}=10^4$~ years, the truncation radius decreases according to \equ{eq:truncation_radius} almost instantly on a timescale $\tau_\mathrm{dyn,~in}$ (panel b).
The stellar rotation period and the corotation radius remain approximately constant as the stellar spin evolves on a larger timescale $\tau_\mathrm{spin}$ (see \fig{fig:timescales}).
Consequently, the fastness parameter decreases from its initial value of 0.72 toward $\approx 0.17$ (or by a factor of $\approx 4$).
As long as the accretion rate remains above $\Dot{M}_\star \gtrsim 10^{-9}~\Msolpyr$, the stellar spin slowly adjusts to the new condition, the star spins up, and $\omega_\mathrm{s}$ increases over the next few million years.
As the accretion rate decreases due to viscous evolution and PE, the truncation radius increases, ultimately resulting in a value of $\omega_\mathrm{S}\approx 1$ as the accretion rate drops toward zero.
During their T~Tauri lifetime, stars at different masses ($M_\star \gtrsim 0.7~\mathrm{M_\odot}$) start to develop a radiative core at different ages, and for all these different stellar masses, observations show a significant scatter in accretion rates.
As a result, numerous combinations of stellar masses, ages, and accretion rates yield different rotation periods in their respective ZTS. 
For both stellar masses, which are shown here at different ages, the effect of a decreasing dipole field strength due to the development of a radiative core is similar.

\begin{figure}
    \centering
         \resizebox{\hsize}{!}{\includegraphics{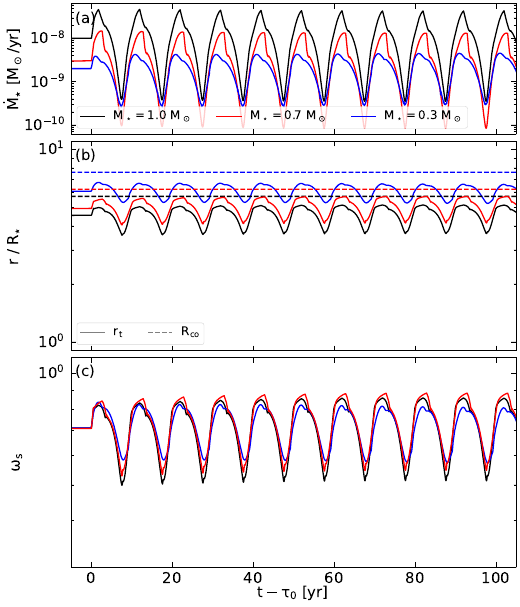}}
    \caption{Evolution of the short model. 
    The panels are the same as in \fig{fig:long}.
    The time reference is the beginning of the magnetic cycles at a simulation time of $\tau_\mathrm{0}=10^4$~years.
    }
    \label{fig:short}
\end{figure}

In the case of stellar magnetic cycles, the corresponding timescale $\tau_\mathrm{cyc}=10$~years is significantly smaller than the stellar or disk evolution timescales.
Thus, we assumed that the stellar radius remains constant over several $\tau_\mathrm{cyc}$.
In contrast to the development of a radiative core, stellar magnetic cycles can also affect smaller stars that remain fully convective during their disk phase.
The evolution of the short model for three different stellar masses, ages, and accretion rates is shown in \fig{fig:short}.
With an increasing magnetic field strength, the truncation radius is pushed away from the star (see \equo{eq:truncation_radius}).
The disk material cannot adjust to the change in the truncation radius as the viscous timescale in the inner disk is significantly larger compared to the cycles timescale ($\tau_\mathrm{\nu,in} /\tau_\mathrm{cyc}\gtrsim100$).
As a consequence, more disk material is located inside $r_\mathrm{t}$ and is accreted on the star, increasing the accretion rate (panel a).
Similarly, as $B_\star$ decreases, the disk material cannot follow the decreasing truncation radius, and the accretion rate also decreases.
We note that for all our simulations the condition $r_\mathrm{t}<R_\mathrm{co}$ or $\omega_\mathrm{S}<1$ is fulfilled.
Should the increase in magnetic field strength push the truncation radius beyond the corotation radius, disk material would be propelled away from the star and the accretion rate would significantly drop \citep[e.g.,][]{Armitage1995, Romanova2004prop}.
The slow back-reaction in the accretion rate ($\tau_\mathrm{\nu,in} > \tau_\mathrm{{cyc}}$) weakens the effect of the magnetic cycles on the ZTS (the change in the accretion rate has an opposite effect on $r_\mathrm{t}$ compared to the variable magnetic field, panel c and \equo{eq:truncation_radius}). 
During one cycle, $\omega_\mathrm{s}$ varies between 0.87 and 0.50 (or by a factor of 1.74).
Although the absolute values vary, the overall behavior of the accretion rate and $\omega_\mathrm{S}$ is comparable for the different stellar masses, ages, and accretion rates.

\section{Discussion}
\label{sec:discussion}

\subsection{Comparison with observations}

\begin{figure}
    \centering
         \resizebox{\hsize}{!}{\includegraphics{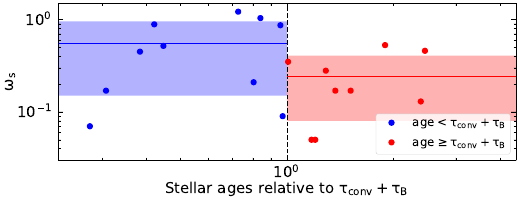}}
    \caption{Comparison of the values of $\omega_\mathrm{S}$ of the young (ages $<\tau_\mathrm{conv}+\tau_\mathrm{B}$) and old (ages $\geq\tau_\mathrm{conv}+\tau_\mathrm{B}$) stars listed in \tab{tab:ages}. 
    The stellar ages are plotted relative to $\tau_\mathrm{conv}+\tau_\mathrm{B}$ (marked by the vertical dashed line).
    The horizontal lines show the respective mean values, and the colored areas mark the respective standard deviations.
    }
    \label{fig:ages_comp}
\end{figure}

Recently, the truncation radius of T~Tauri stars has been derived from observational data in combination with the stellar rotation period and the corotation radius for a few dozen stars \citep[][]{Thanathibodee2023, Pittman2025rot}.
In both cases, a significant number of stars show values $\omega_\mathrm{S}$ smaller than expected from stellar spin models \citep[e.g.,][]{Ireland2022, Serna2024, Zhu2025}.
Based on the results presented in \cite{Pittman2025rot}, we wanted to compare the $\omega_\mathrm{S}$ values of fully convective stars and stars that have started to develop a radiative core.
For this comparison, we selected stars with stellar masses of $M_\star \geq 0.7~\mathrm{M_\odot}$ (see \tab{tab:ages}). 
For these stellar masses, the radiative core is expected to develop within the first 5~Myr of their lifetimes.
For each star, we compared the stellar age taken from the literature with the end of the fully convective phase $\tau_\mathrm{conv}$ based on the stellar evolution models of \citet[see also \citealt{Gregory2012}]{Tognelli11}.
\begin{equation}\label{equ:tau_conv}
    \tau_\mathrm{conv}~\mathrm{[Myr]} \approx \left( \frac{1.448}{M_\star / M_\odot}\right)^{2.101} \, .
\end{equation}
We note that a minimum value of $\omega_\mathrm{S}$ should be reached at around $\tau_\mathrm{B}= 0.5$~Myr after the radiative core starts to develop (see \fig{fig:long}).
For a $1.0~\mathrm{M_\odot}$ star, this results in a stellar age of $\tau_\mathrm{conv}+\tau_\mathrm{B}\approx2.6$~Myr and for a $0.7~\mathrm{M_\odot}$ star, in an age of $\tau_\mathrm{conv}+\tau_\mathrm{B}\approx5.1$~Myr according to \equo{equ:tau_conv}.
We distinguish between stars with ages older ("old" stars) and younger ("young" stars) than $\tau_\mathrm{conv}+\tau_\mathrm{B}$ and compare the values of $\omega_\mathrm{S}$ in \fig{fig:ages_comp}.
For young stars, the mean fastness parameter is $\omega_\mathrm{S}= 0.55\pm0.40$ and for old stars, $\omega_\mathrm{S}= 0.24\pm0.16$.
The values for young stars are consistent with the ZTS for a spin model including an APSW \citep[see][]{Gehrig2025Letter}.
The values for old stars match the simulation results shown in \fig{fig:long}b.

While this initial comparison suggests an age trend in $\omega_\mathrm{S}$, we note that some caution remains warranted.
First of all, the small number of available observations and the resulting uncertainties require more measurements to draw a clear conclusion.
The two stars with the largest values of $\omega_\mathrm{S}$, MY~Lup and CV~Cha, are in the regime $\tau_\mathrm{conv}<\mathrm{age}<\tau_\mathrm{conv}+\tau_\mathrm{B}$.
According to \fig{fig:long}, some initial effects of the decreasing dipole field component on $\omega_\mathrm{S}$ should already be noticeable.
Although this dataset is not fully consistent with the results presented in this work, the observational uncertainties should be taken into account.
The inner disk radius and, in turn, $\omega_\mathrm{S}$ are subjected to short-term variability, for example, due to variability in the accretion rate \citep[e.g.,][]{Zsidi2022, Robinson2022}.
In the case of MY~Lup, the value of $\omega_\mathrm{S}$ varies from 0.55 to 1.42 during the observations \citep[][]{Pittman2025rot}.
Further, mass and age estimates can vary significantly depending on the observation method itself \citep[e.g.,][]{Flores2022}.
For example, the age of CV~Cha is reported to be 1.3-1.5~Myr \citep[from which the mean value was used in \tab{tab:ages};][]{Ginski2024} and $3.0-4.5$~Myr \citep[][]{Gregory2012}.
Another potential source of uncertainty in stellar age estimates is the disk itself.
A large inclination (as in the case of MY~Lup) and flaring can extinct the host star, leading to an overestimation of the stellar age \citep[e.g.,][]{Andrews2018}.
As a result, the age of MY~Lup has been corrected from 16.6~Myr, as reported in \cite{Frasca2017}, to 2~Myr \citep[][]{Long2022}.
In addition, the effects of binaries and magnetic stellar evolution models \citep[e.g.,][]{Feiden2016} are not included in this preliminary comparison.

We further note that the value of $\omega_\mathrm{S}$ should increase again due to decreasing accretion rates toward the end of the disk's lifetime (see \fig{fig:long}).
For all but two stars from the sample of \cite{Pittman2025rot}, the accretion rate shows values of $\Dot{M}_\star > 10^{-9}~\Msolpyr$ even at older ages, and small values of $\omega_\mathrm{S}$ are expected after the radiative core has formed, matching our results.
The accretion rates provided by \cite{Thanathibodee2023}, on the other hand, range at $\Dot{M}_\star \sim 10^{-10}~\Msolpyr$ and they find values of $\omega_\mathrm{S}\lesssim 0.45$ for most stars in their study, contradicting our results.
As discussed in \cite{Thanathibodee2023}, the mechanisms that shape the final stages of disk accretion in T~Tauri stars have to be studied in future work.
However, analyzing the stellar spin state (e.g., by means of the fastness parameter) at different stellar evolutionary stages (e.g., $\mathrm{age} <\tau_\mathrm{conv}$, $\mathrm{age}\approx \tau_\mathrm{conv}$, and $\mathrm{age} >\tau_\mathrm{conv}$) can advance our understanding of the star--disk interaction.
One mission that could provide the fastness parameter for many stars is the Early eVolution Explorer (EVE) satellite \citep[][]{MacGregorEVE}.
With a combined derivation of the truncation radius, the accretion rate, and the stellar rotation period, EVE can increase the available data by a factor of $\gtrsim 15$.

Comparing our results on magnetic cycles with observational data proves to be no less challenging.
While variation in the dipole field strengths on the timescales of years have been known for a few stars \citep[e.g.,][]{Johnstone14, Finociety2023V1298, Gaidos2024}, the first magnetic polarity reversal of a T Tauri star has been reported recently \citep[for the star DO~Tau, see][]{Donati2026}.
A suitable comparison with our results requires simultaneous measurements of the dipole field strength, the truncation radius \citep[that can be calculated with the accretion rate and the dipole field strength][]{bessolaz08, hartmann16}, and the stellar rotation period (setting the corotation radius).
The SpectroPolarimetre InfraRouge (SPIRou) Legacy Survey does provide these measurements \citep[e.g.,][]{Donati2020SPIROU, Donati2024a, Donati2026}.
In the course of the survey, young, accreting stars show variations of the dipole field strengths on timescales of several years.
The dipole strength of V1298~Tau decreases from 245~G to 85~G (a factor of almost 3) during one year \citep[][]{Finociety2023V1298} and during the polarity reversal of DO~Tau, the dipole field evolves from $-190$~G to 320~G \citep[][]{Donati2026}.
The absolute change of the dipole field in these stars is relatively small ($\sim 100$~G) compared to the present study and effects on the accretion rate are not clearly measurable.
Larger dipole variations are reported for V2129~Oph from 280~G in 2005 to 970~G in 2009 \citep[][]{Johnstone14}.
Unfortunately, the accretion rates have not been measured simultaneously and the effect of the different dipole field strengths on the accretion rate, the truncation radius, and the stellar spin state cannot be estimated.
Future results by surveys like SPIRou will provide valuable data for constraining stellar magnetic cycles and their effects on the stellar spin evolution.

\begin{table}[ht]
\centering
\caption{
Mean stellar parameters from \cite{Pittman2025rot}, including stellar ages taken from the literature.
}  
\label{tab:ages} 
\renewcommand{\arraystretch}{1.125}
\begin{tabular}{lccccc}         
\hline\hline 

Name & Mass & $\omega_\mathrm{S}$ & Age & Age & Age\\
&$\mathrm{[M_\odot]}$&&[Myr]& class & ref\\
\hline

MY Lup  &1.06 &1.22&2.0&y&1\\
CV Cha  &1.40&1.04&1.4&y&2\\
CVSO 58 &0.81&0.89&2.0&y&3\\
Sz 19   &2.37&0.87&0.8&y&2\\
CS Cha  &1.22&0.53&4.0&o&2\\
Sz 68   &1.83&0.52&0.5&y&4\\
RX J1842.9&1.04&0.46&10&o&5\\
-3532   &&&&&\\
TX Ori  &1.09&0.45&1.0&y&6\\
UX Tau A        &1.50&0.35&1.5&o&7\\
V510 Ori        &0.76&0.28&7.0&o&8\\
SZ Cha  &1.22&0.21&1.7&y&2\\
AA Tau  &0.80&0.17&1.5&y&9\\
RX J1852.3&0.82&0.17&10&o&5\\
-3700   &&&&&\\
V505 Ori&0.81&0.17&6.5&o&8\\
CVSO 146 &      0.86&0.13&10&o&6\\
RECX 11 &       0.78&0.09&5.0&y&10\\
SY Cha  &0.70&0.07&1.8&y&2\\
RY Lup  &1.40&0.05&2.0&o&11\\
J160830.7&      1.38&0.05&2.0&o&12\\
-382827&&&&&\\

\hline\hline  
\end{tabular}
\newline
\vspace{-0.2cm}
\begin{flushleft}
\tablefoot{
The column "Age class" corresponds to young (y) and old (o) stars with respect to $\tau_\mathrm{conv} +\tau_\mathrm{B}$.
Age references: (1) \cite{Long2022}, (2) \cite{Ginski2024}, (3) \cite{Pittman2022}, (4) \cite{Frasca2017}, (5) \cite{Roccatagliata2009}, (6) \cite{Valegard2024}, (7) \cite{Menard2020}, (8) \cite{Mauco2016}, (9) \cite{Donati2010}, (10) \cite{Rugel2018}, (11) \cite{GRAVITY_Collaboration2020}, and (12) \cite{Biazzo2017}.}\\
\end{flushleft}
\end{table}

\subsection{Other factors affecting the stellar spin state}

During the evolution of a T~Tauri star, other factors can affect the stellar spin state through the truncation radius or the corotation radius.
In terms of the truncation radius, the most volatile contributing factors are the magnetic field strength and the accretion rate (see \equo{eq:truncation_radius}).
The accretion rate can be variable on timescales of days to years \citep[see the review by][]{Fischer2023}.
Observations show a routine variability of $<1-2$~mag over days to years \citep[matching the dynamic and viscous timescale of the inner disk; e.g.,][]{Cody2017}.
In addition, outburst events can increase the accretion rate by up to 6~mag and can last for years to decades \citep[e.g., Fu~Ori outbursts;][]{Kospal2011}.
While we have observed a back-reaction of the accretion rate on the stellar magnetic cycles, there are other reasons for a variable $\Dot{M}_\star$ that are independent of $B_\star$ \citep[e.g.,][]{Fischer2023}.
According to \equo{eq:truncation_radius}, a one-order-of-magnitude increase in the accretion rate will push the truncation radius inward by a factor of 0.52, corresponding to a decrease in $\omega_\mathrm{s}$ by 63\%.
Together with the effect of a stellar magnetic cycle, the value of $\omega_\mathrm{s}$ can be decreased even further.
The potential combination of these effects could lead to a significant deviation from the ZTS.

Another factor that can affect the stellar spin state of observed stars is the assumed spin model itself.
The ZTS value of $\omega_\mathrm{s}$ varies with different spin models, as the torque contributions acting on the star are subject to change with theoretical and observational advances \citep[e.g.,][]{Ireland21, Ireland2022, Takasao2025, Zhu2025}.
In their 3D model, for example, \cite{Takasao2025} found that MEs might play only a minor role in the stellar spin evolution, whereas CDWs can potentially remove significant amounts of disk and stellar AM (comparable to a strong APSW).
An additional torque contribution that removes stellar AM, similar to an APSW, can result in smaller $\omega_\mathrm{S}$ values in the ZTS \citep[][]{Gehrig2025Letter}.
Unfortunately, these recent models are restricted to a limited parameter space due to high computational costs \citep[e.g.,][]{Takasao2022, Zhu2025}, and generalizing the results might not be straightforward.

\begin{figure}
    \centering
         \resizebox{\hsize}{!}{\includegraphics{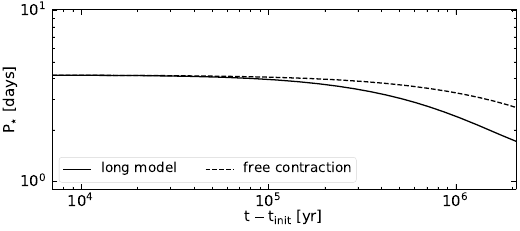}}
    \caption{Evolution of the stellar rotation period of the long model for a stellar mass of $1.0~\mathrm{M_\odot}$.
    The rotation period of a contracting star according to \equo{eq:contraction} without any external torque contributions is shown for comparison. 
    }
    \label{fig:Pstar}
\end{figure}

This apparent incompleteness of the stellar spin model can be illustrated by the evolution of the stellar rotation period.
In \fig{fig:Pstar} we show the rotation period of the long model for a stellar mass of $1.0~\Msol$ (see also \fig{fig:long}).
For comparison, the stellar rotation period of a contracting star according to \equo{eq:contraction} without any external torque contributions is highlighted.
With a decreasing stellar magnetic field strength and low values of $\omega_\mathrm{S}$, a spin-up torque is exerted onto the star in addition to its contraction, leading to fast rotation periods.
This spin-up contradicts the observed constancy of rotational distributions \citep[e.g.,][]{Smith2023}, which indicates that certain spin-down contributions are not yet completely covered by the stellar spin model, as stated above, for example, the effects of winds in the innermost disk region.

\subsection{Dependence on timescales}\label{sec:timescales}

\begin{figure}
    \centering
         \resizebox{\hsize}{!}{\includegraphics{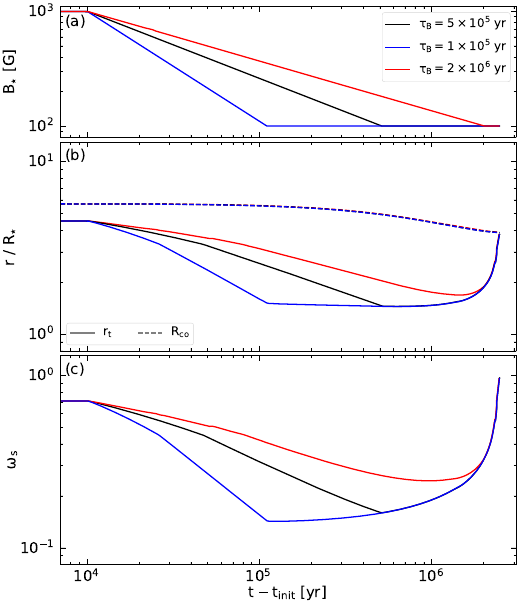}}
    \caption{Effect of different values of $\tau_\mathrm{B}$ on the long model.
    The stellar mass is $1.0~\mathrm{M_\odot}$.
    Panel (a): Dipole field strength, $B_\star$.\ Panels (b) and (c): Same as in \fig{fig:long}.
    }
    \label{fig:fs}
\end{figure}

We have shown the effects of a variable $B_\star$ based on two different scenarios acting on two different timescales.
These timescales, however, might vary according to different stellar masses or ages.
First, we show how the long model reacts if we change $\tau_\mathrm{B}$ to $10^5$~years and $2\times10^6$~years.
The stellar spin reacts to a changing magnetic field strength on timescales of $\tau_\mathrm{spin}\gtrsim 2-3$~Myr (see \fig{fig:long}).
If the timescale on which the magnetic field changes is significantly smaller, $\tau_\mathrm{B}\ll \tau_\mathrm{spin}$, the stellar spin cannot adjust to the new parameters and the relative effect on $\omega_\mathrm{S}$ is large (blue lines in \fig{fig:fs}).
For increasing values of $\tau_\mathrm{B}\sim \tau_\mathrm{spin}$, the stellar spin can be adjusted to the changing parameter, and the relative effect on $\omega_\mathrm{S}$ is weaker (red lines in \fig{fig:fs}).
If $\tau_\mathrm{B}$ were much larger than $\tau_\mathrm{spin}$, we would not expect a significant shift away from the ZTS.
In addition to the adjustment of the stellar spin to the new magnetic field strength, the decrease in the accretion rate due to the disk evolution and PE can limit the effect on $\omega_\mathrm{s}$ in the case of large values of $\tau_\mathrm{B}$.

\begin{figure}
    \centering
         \resizebox{\hsize}{!}{\includegraphics{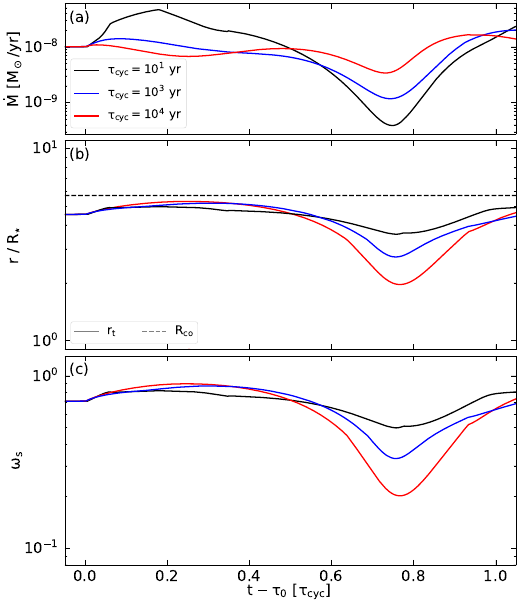}}
    \caption{Effect of different values of $\tau_\mathrm{cyc}$ on the short model.
    The stellar mass is $1.0~\mathrm{M_\odot}$.
    The time axis is normalized to one cycle, starting at $\tau_\mathrm{0}$.
    The panels are the same as in \fig{fig:long}.
    }
    \label{fig:short_tau}
\end{figure}

For variations in magnetic field strength on shorter timescales, we show how different values of $\tau_\mathrm{cyc}$ affect the evolution of $\omega_\mathrm{s}$ (see \fig{fig:short_tau}).
As shown in \fig{fig:short}, the disk material is slow to adjust to variations shorter than the viscous timescale in the inner disk ($\tau_\mathrm{cyc}<\tau_\mathrm{\nu, in}\approx 1000$~years, black lines).
As a result, the accretion rate is affected by the magnetic cycle (the variable accretion rate has an opposite effect on $r_\mathrm{t}$ compared to the variable magnetic field), and the effect on $\omega_\mathrm{S}$ is weakened.
For increasing values of $\tau_\mathrm{cyc}=\tau_\mathrm{\nu, in}$, the disk material can adjust to the magnetic field variation, and enough disk material can be redistributed within the inner disk. 
As a result, the effect of a variable magnetic field on the accretion rate becomes smaller for increasing values of $\tau_\mathrm{cyc}$, and the accretion rate remains closer to its initial value.
Thus, the $\omega_\mathrm{S}$ is shifted further away from its ZTS (blue lines).
For even larger values $\tau_\mathrm{\nu, in} <\tau_\mathrm{cyc}=10^4~\mathrm{years}$, $\omega_\mathrm{s}$ is pushed toward values of $\sim 0.2$ (red lines)\footnote{
We note that large values of $\tau_\mathrm{cyc}\gtrsim100$~years are yet to be confirmed for T~Tauri stars.
Based on ice-core reconstructions, \cite{Vonmoos2006} found variations in solar activity ranging around $10^3-10^4$~years that might also be applicable for younger stars.
}.
While $\omega_\mathrm{s}$ is shifted significantly toward smaller values as a result of a decreasing field strength during a magnetic cycle, the shift toward larger values is weaker.
When the truncation radius is pushed outward by an increasing magnetic field, the disk pressure increases close to the corotation radius \citep[e.g.,][]{Steiner21, Ireland2022}, preventing a further outward shift.
Summarizing, the effect of short-term magnetic variations on $\omega_\mathrm{S}$ depends on the relation between $\tau_\mathrm{cyc}$ and the viscous timescale of the inner disk $\tau_\mathrm{\nu, in}$.
A shift of the disk's inner radius toward the star will also increase the temperature structure of the innermost disk regions.
If the temperature increase is sufficient to ionize enough disk material, magneto-rotational instabilities can be triggered, causing further variability.
These effects, however, are beyond the scope of this work and might be the subject of future studies.

As mentioned in Sect. \ref{sec:model}, the current model does not include the effects of magnetohydrodynamic-driven disk winds.
There is growing theoretical and observational evidence that disk winds are essential for the evolution of PPDs \citep[e.g.,][]{Zhu2018, Jacquemin-Ide2021, Pascucci2023, vanDishoeck2025, Zhu2025, Feeney-Johansson2026}.
Within the context of the present work, we want to discuss the effect of such disk winds on our results.
The effect of a variation in the stellar magnetic field on the spin state depends on how fast the disk can react to changes in field strength.
In a disk evolution scenario dominated by magnetohydrodynamic winds, the accretion timescale, $\tau_\mathrm{acc}$, is expected to be shorter compared to the viscous timescale.
Models yield estimates of around $\tau_\mathrm{acc}\sim 100\times \tau_\mathrm{dyn}$ \citep[e.g.,][]{Jacquemin-Ide2021}.
In the outer disk regions, this is comparable to the viscous timescale \citep[see also][]{Tabone2021}, and our results should also apply to the wind-driven scenario.
In the inner disk, on the other hand, the accretion timescale is significantly shorter than the viscous timescale, $\tau_\mathrm{acc} \lesssim \tau_\mathrm{cyc}\ll \tau_\mathrm{\nu,~in}$ (see also \fig{fig:timescales}).
As a result, material in the inner disk would react more quickly, and the stellar spin state should become more sensitive to variations in the stellar magnetic field strength (this would be comparable to larger values of $\tau_\mathrm{cyc}$ in \fig{fig:short_tau}).
The precise effects of disk winds on the stellar spin state will be the subject of future studies of this model, extended with magnetohydrodynamic-driven winds \citep[with the fundamentals already included in][]{Steiner2025}.

\subsection{Further implications for the inner disk}
\label{sec:reversal}

In this work, we did not include the large-scale disk magnetic field \citep[for recent developments we refer to][]{Steiner2025} or the effects resulting from the reversal of the magnetic field's polarity (see Sect. \ref{sec:Bvar}). 
However, the stellar magnetic field can dominate the large-scale magnetic field topology in the inner disk region \citep[see][]{Steiner2025}. 
Recent 3D simulations of the star-disk connection have shown that only a very limited region of the disk is directly connected to the star \citep[$r\lesssim R_\mathrm{co}$; e.g.,][]{Takasao2022, Zhu2025}. 
Stellar field lines threading the disk at larger radii open up and are not able to transfer AM between the star and the disk. 
However, a changing polarity of the stellar magnetic field has implications on the resulting magnetic field strength and wind launching mechanisms \citep[e.g.,][]{Ferreira06} in the very inner disk:
\begin{enumerate}[(i)]
    \item In the anti-aligned case, magnetic flux, transported inward by magnetic advection, builds up close to the inner disk region, where the stellar magnetic field is strong enough to support closed field lines. This leads to an increased magnetic field strength in the vicinity of the magnetic truncation radius. 
    In this configuration, an out-bursting, reoccurring outflow (for example, MEs) can transport AM between the star and the disk \citep[e.g.,][]{Romanova09, Zanni13}.
    \item In the aligned case, the magnetic flux can merge/reconnect with the stellar magnetic field.
    At the point, where the stellar field "cancels out" the disk field (the so-called X-point), open field lines can connect to the closed stellar field lines and remove AM \citep[e.g.,][]{Ferreira00}.
\end{enumerate}
An oscillating polarity therefore likely leads to a time-dependent change in magnetic field strength and field topology close to the inner disk rim. 
This has an impact on the possibility and strength outflows originating from the innermost disk region \citep[e.g.,][]{Lesur2023}.
Thus, periodically changing stellar fields could be a possible source of time-dependent disk winds. 
Recent observations of disk outflows and jets show "knotty" jets, which could be caused by a change in the magnetic field polarity \citep[e.g.,][]{vorobyov2018knottyjets,bajaj2025}.
To study the effects of changing magnetic field polarity on the inner disk and winds, future models should combine: (i) the innermost disk region including the magnetic star-disk connection, (ii) a description of the large-scale magnetic field, and (iii) a wind model, describing the launch of magneto-centrifugal and magneto-thermal winds.

\section{Summary}
\label{sec:summary}

In this work we show that the temporal variation of the stellar dipole field strength ($B_\star$) can have a distinct effect on the stellar spin state, described by the fastness parameter ($\omega_\mathrm{s}$).
Observations have shown significantly lower values of $\omega_\mathrm{s}$ than expected from stellar spin models.
Two potential scenarios of a variable magnetic field have been studied.
First, the decrease in $B_\star$ due to the development of a radiative core on a timescale of $\tau_\mathrm{B} = 0.1-2.0$~Myr (long model).
Second, the variability of $B_\star$ due to stellar magnetic cycles with $\tau_\mathrm{cyc}=10^1-10^4$~years (short model).
The strength of these variations depends on the relation between the timescale of the changing magnetic field ($\tau_\mathrm{B}$ or $\tau_\mathrm{cyc}$), the spin-up timescale ($\tau_\mathrm{spin}$), and the viscous timescale of the accretion disk ($\tau_\mathrm{\nu}$).

In the case of $\tau_\mathrm{B}\ll\tau_\mathrm{spin}\sim 2-3$~Myr, the stellar spin is slow to adjust to the changing parameters, and $\omega_\mathrm{S}$ is shifted out of its ZTS.
If the variations occur on long timescales $\tau_\mathrm{B}\sim\tau_\mathrm{spin}$, the stellar spin can adjust to the change and remain closer to its ZTS.
On shorter timescales, the effect of magnetic cycles on $\omega_\mathrm{S}$ is weaker if they occur on shorter timescales than the viscous timescale of the inner disk $\tau_\mathrm{cyc}\ll \tau_\mathrm{\nu, in}\sim 10^3$~years.
The inner disk material is slow to adjust to the changing parameters, resulting in a back-reaction in the accretion rate.
For $\tau_\mathrm{cyc}\gg \tau_\mathrm{\nu, in}$, this back-reaction is less pronounced and the effect on $\omega_\mathrm{S}$ increases.
Our results can explain at least part of the observed low values $\omega_\mathrm{s}$.
Another possibility that could explain the differences between expected and measured $\omega_\mathrm{S}$ values is outflows and winds from the inner disk region.
These winds can transport significant amounts of AM and affect the stellar spin state.
In the next iteration of our star-disk model, we focus on these (time-dependent) disk winds in different configurations.
We encourage further theoretical and observational work that connects accretion, stellar rotation, and magnetic properties in T~Tauri stars.

\begin{acknowledgements}
The authors thank the anonymous referees for constructive feedback and insightful comments that improved the clarity of this work.
Furthermore, we thank Eric Gaidos for his valuable comments during the course of the work.
All figures were created using the Python packages Matplotlib \citep[][]{Hunter2007Matplotlib} and Numpy \citep[][]{Harris2020Numpy}.
 
\end{acknowledgements}


\bibliographystyle{aa}
\bibliography{main}


\label{LastPage}
\end{document}